\documentclass[footinbib,superscriptaddress,preprintnumbers,amsmath,amssymb]{revtex4-1}

\usepackage{graphicx}
\usepackage{subfigure}
\usepackage{epsfig}
\usepackage{dcolumn}
\usepackage{bm}

\begin{document}

\title{Full-counting statistics of energy transport of molecular junctions in the polaronic regime}

\author{Gaomin Tang}
\affiliation{Department of Physics and the Center of Theoretical and Computational Physics, The University of Hong Kong, Hong Kong, China and The University of Hong Kong Shenzhen Institute of Research and Innovation, Shenzhen, China}
\author{Zhizhou Yu}
\affiliation{School of Physics and Technology, Nanjing Normal University, Nanjing 210023, China}
\affiliation{Department of Physics and the Center of Theoretical and Computational Physics, The University of Hong Kong, Hong Kong, China and The University of Hong Kong Shenzhen Institute of Research and Innovation, Shenzhen, China}
\author{Jian Wang}
\email{jianwang@hku.hk}
\affiliation{Department of Physics and the Center of Theoretical and Computational Physics, The University of Hong Kong, Hong Kong, China and The University of Hong Kong Shenzhen Institute of Research and Innovation, Shenzhen, China}

\date{\today}

\begin{abstract}
We investigate the full-counting statistics (FCS) of energy transport carried by electrons in molecular junctions for the Anderson-Holstein model in the polaronic regime. Using two-time quantum measurement scheme, generating function (GF) for the energy transport is derived and expressed as a Fredholm determinant in terms of Keldysh nonequilibrium Green's function in the time domain. Dressed tunneling approximation is used in decoupling the phonon cloud operator in the polaronic regime. This formalism enables us to analyze the time evolution of energy transport dynamics after a sudden switch-on of the coupling between the dot and the leads towards the stationary state. The steady state energy current cumulant GF in the long time limit is obtained in the energy domain as well. Universal relations for steady state energy current FCS are derived under finite temperature gradient with zero bias and this enables us to express the equilibrium energy current cumulant by a linear combination of lower order cumulants. Behaviors of energy current cumulants in steady state under temperature gradient and external bias are numerically studied and explained. Transient dynamics of energy current cumulants is numerically calculated and analyzed. The universal scaling of normalized transient energy cumulants is found under both temperature gradient and external bias. 
\end{abstract}

\maketitle

\section{Introduction}\label{sec1}
Rapid experimental development in the field of nanotechnology makes fabrication of the single-molecule junctions possible\cite{molecule1, molecule2}, which could push the limit of Moore's law further. In the electronic quantum transport though nano-devices, the electron-phonon coupling plays an important role. One of the mechanisms that induces electron-phonon coupling is due to
the charging of the molecule leading to elastic mechanical deformations. This in turn causes an interaction between electronic and the quantized mechanical degrees of freedom giving rise to the electron-phonon coupling. A variety of intriguing transport properties, such as phonon-assisted current steps and Franck-Condon blockade \cite{FC_blockade} have been found in the polaronic regime  \cite{ep_polaron1, ep_polaron2} when this kind of electron-phonon coupling in molecular junctions is strong.
Theoretically, these phenomena could be understood using a quantum dot described by the Anderson-Holstein model \cite{Holstein, Mahan} coupled to two electrodes.

	To understand quantum transport in the polaronic regime, many methods have been used, such as the master equation method\cite{ME1, ME2, ME3, ME4}, diagrammatic quantum Monte Carlo method \cite{diagMC}, numerical renormalization group method\cite{NRG}, as well as the nonequilibrium Green's function (NEGF) technique\cite{yeyati_FCS} that is particular useful in describing time dependent non-equilibrium processes.
Perturbation method is applicable when the electron-phonon coupling strength is weak\cite{perturb1,perturb2,perturb3} and it fails in the strong electron-phonon coupling system.
Other approximation has to be made in order to deal with the strong and intrinsically nonlinear electron-phonon interaction in the Anderson-Holstein model.
In order to decouple the phonon cloud operator in the polaronic regime, dressed tunneling approximation (DTA), in which the leads' self-energies are dressed with the polaronic cloud, has been proposed to eliminate the noticeable pathological features of the single particle approximation (SPA) at low frequencies and polaron tunneling approximation (PTA) at high frequencies \cite{DTA, BDong, BDong_SC, yeyati_FCS}.
	
	It is known that quantum transport is determined in nature by stochastic process which could be characterized by the corresponding distribution function \cite{Blanter}.
The study of Full-counting statistics (FCS) pioneered by Levitov and Lesovik \cite{Levitov1, Levitov2, Levitov3} could give us a full scenery of probability distribution of electron and energy transport \cite{BDong, BDong_SC, yeyati_FCS, Klich1, Nazarov, RMP, wavepacket, JS1, JS2, JS3, gm1, gm2, gm3, Yuan, gm6}. The key in FCS is to obtain the generating function (GF) which is actually the Fourier transform of probability distribution of the related physical quantity. Using the NEGF technique \cite{Keldysh, Haug, Ka1, Ka2} and the path integral method under the two-time quantum measurement scheme \cite{RMP, twotime1, twotime2, twotime3}, GF was formulated as a Fredholm determinant in the time domain for both phonon \cite{JS1, JS2, JS3} and electron \cite{RMP, gm1, gm2, gm3, Yuan} transport. This formalism enables one to study the transport properties in the transient regime providing more information on the short time dynamics \cite{gm1}. Recently transient dynamics of particle current transport in the molecular junctions has been studied by Schmidt \textit{et al.} \cite{Schmidt1, Schmidt2} in the case of weak and strong electron-phonon couplings
and has been reported by Maier \textit{et al.} using PTA \cite{Schmidt3} and by Souto \textit{et al.} using DTA \cite{yeyati_FCS} in the polaronic regime.
	
	The transport study of energy flow in the nonequilibrium system could reveal information on how energy is dissipated and its correlation for electronic devices and can be investigated theoretically by Landauer-B\"{u}ttiker type of formalism for noninteracting systems \cite{LB-energy1, LB-energy2, LB-energy3}. Energy transport in trapped ion chains has been measured experimentally by Ramm \textit{et al.} \cite{energy_measure}. The heat current $I^h_\alpha$ in the $\alpha$ lead is related to the energy current $I^E_\alpha$ by the expression $I^h_\alpha = I^E_\alpha-\mu_\alpha I_\alpha$ with the particle current $I_\alpha$ and the chemical potential $\mu_\alpha$ in the $\alpha$ lead, and this quantity is quite important in characterizing the efficiency of thermoelectric devices \cite{themo}.
So far, FCS of energy transfer mostly focuses on phonon transport both in the transient regime and steady states \cite{JS1, JS2, JS3} and less attention has been paid to the FCS of
energy transfer carried by electrons in the electronic transport problems. In our previous work, we investigated the transient FCS of energy transfer in the non-interacting system \cite{gm3}.
It would be important and interesting to study of FCS of energy transport carried by electrons of molecular junctions with electron-phonon coupling in the polaronic regime for both transient dynamics and steady states, and this is the purpose of this work.

	In this paper, FCS of energy transport carried by electrons in molecular junctions for the Anderson-Holstein model in the polaronic regime is investigated both in the steady states and transient regime. Within the DTA, GF for the energy current is derived from the equation of motion and could be expressed as a Fredholm determinant in the time domain using NEGF. Numerical calculation is performed which allows us to analyze the time evolution of the energy flow towards the steady state for a sudden switch-on of the coupling between the quantum dot and the leads. The cumulant GF of energy current in the steady state is obtained analytically in the energy domain. Universal relations for cumulants of energy current under finite temperature gradient with zero voltage bias are established. In addition, we also calculate and analyze steady state solution for various order of cumulants (from the first to the fourth order) under temperature gradient or external bias.

	The rest of the paper is organized as follows. In Sec. II, the model Hamiltonian of a molecular junction is introduced and GF of energy flow in the transient regime is determined in terms of NEGF in the time domain. Sec. III is devoted to the steady state investigation of FCS of energy current, both theoretically and numerically. In Sec. IV, transient dynamics of energy current is investigated under a sudden switching-on of external bias. Finally, a brief conclusion is drawn in Sec. V.

\section{Model and basic theoretical formalism}
\begin{figure}
  \includegraphics[width=3.4in]{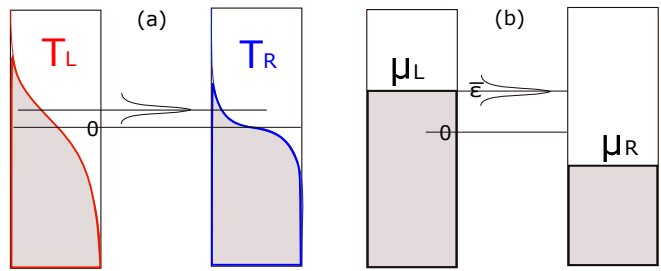} \\
  \caption{Sketch of a QD coupled the left and right lead under (a) temperature gradient $T_L>T_R$ with zero chemical potentials in both leads and (b) external bias $\Delta \mu$ with $\mu_{L(R)}=\pm \Delta\mu/2$ under zero temperature. }
  \label{fig1}
\end{figure}

	Considering only the lowest electronic orbital, the single-molecule is simplified as a single electronic level of a quantum dot (QD) being coupled to localized vibrational mode, which is the simplest spinless Anderson-Holstein model \cite{Egger}. The QD then is coupled to the left and right electrode so that the system is driven to a nonequilibrium state when the external bias or temperature gradient is applied (Fig.~\ref{fig1}). The corresponding Hamiltonian reads as
\begin{equation}
H = H_{S} + H_L + H_R + H_T
\end{equation}
with the Hamiltonian of the central dot (in natural units, $\hbar = k_B =e = m_e =1$)
\begin{equation}
H_{S} = \epsilon_0 d^\dag d + \omega_0 a^\dag a + t_{ep}(a^\dag +a) d^\dag d,
\end{equation}
where $\epsilon_0$ is the bare electronic energy level, and $\omega_0$ is the frequency of the localized vibron. $d^\dag$ ($a^\dag$) denotes the electron (phonon) creation operator in the QD. The localized vibron modulates the QD with the electron-phonon coupling constant $t_{ep}$. The Hamiltonians of the leads is given in a compact form
\begin{equation}
H_\alpha = \sum_{x\in k\alpha} \epsilon_x c_x^\dag c_x ,
\end{equation}
where the indices $k\alpha = kL, kR$ are used to label the different states in the left and right leads. $H_T$ is the Hamiltonian describing the coupling between the dot and the leads with the tunneling amplitudes $t_{k\alpha}$,
\begin{equation}
H_T = H_{LS} + H_{RS}
= \sum_{k\alpha} (t_{k\alpha} c_{k\alpha}^\dag d + t_{k\alpha}^* d^\dag c_{k\alpha} ) .
\end{equation}
The tunneling rate (linewidth function) of lead $\alpha$ is assumed to bear the Lorentzian form and could be expressed as
\begin{equation} \label{linewidth}
{\bf \Gamma}_\alpha(\omega) ={\rm Im} \sum_k \frac{|t_{k\alpha}|^2}{\omega-\epsilon_{k\alpha}-i0^+} = \frac{\Gamma_\alpha W^2}{\omega^2+W^2} ,
\end{equation}
with the linewidth amplitude $\Gamma_\alpha$ and bandwidth $W$, and one can denote $\Gamma = \Gamma_L+\Gamma_R$.

	The electron-vibron coupling term can be eliminated by applying the Lang-Firsov unitary transformation \cite{Lang-Firsov} given by
\begin{equation}
 \bar{H} = S H S^\dag, \ \ S=e^{g d^\dag d (a^\dag -a)},\ \ g = \frac{t_{ep}}{\omega_0},
\end{equation}
which leads to
\begin{equation}
\bar{H}_{S} = \bar{\epsilon} d^\dag d + \omega_0 a^\dag a ,
\end{equation}
where the bare QD electron energy is changed to $\bar{\epsilon}=\epsilon_0 - g^2\omega_0$. The tunneling Hamiltonian is transformed as
\begin{equation}
\bar{H}_T = \sum_{k\alpha } (t_{k\alpha} c_{k\alpha}^\dag X d + t_{k\alpha}^* d^\dag X^\dag c_{k\alpha})
\end{equation}
with the phonon cloud operator $X=\exp[g(a-a^\dag)]$, while Hamiltonians of isolated leads remain unchanged.

\begin{figure}
  \includegraphics[width=3.2in]{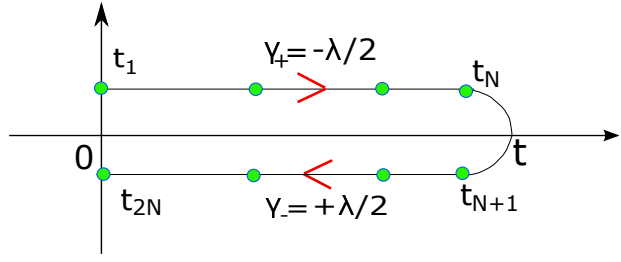} \\
  \caption{ Keldysh contour starting from $t=0$. $\gamma = \pm \lambda/2$ depends on the branch of the contour it resides. }
  \label{fig2}
\end{figure}

	In the present work we study the transient dynamics in which the interaction between the leads and the QD is suddenly turned on at $t=0$ and afterwards the system evolves to the steady states. The turning on process could be facilitated by a quantum point contact which is controlled by a gate voltage. The initial density matrix of the whole system at $t=0$ is the direct product of each subsystem and expressed by $\rho(0) = \rho_L \otimes \rho_S \otimes \rho_R$.
The statistical behaviors of the energy current in a specific lead are all encoded in the probability distribution $P(\Delta \epsilon, t)$ of the transferred energy carried by electrons $\Delta \epsilon = \epsilon_t - \epsilon_0$ between an initial time $t=0$ and a later time $t$.
The GF $Z(\lambda, t)$ with the counting field $\lambda$ is defined as,
\begin{equation}\label{Z}
  Z(\lambda,t) \equiv \langle e^{i\lambda \Delta \epsilon}\rangle = \int P(\Delta \epsilon, t) e^{i\lambda \Delta \epsilon} d\Delta \epsilon.
\end{equation}
The $k$th cumulant of transferred energy $\langle \langle(\Delta \epsilon)^k\rangle\rangle$ could be calculated by taking the $k$th derivative of cumulant generating function (CGF) which is $\ln Z(\lambda)$ with respect to $i\lambda$,
\begin{equation} \label{jth}
C_k(t) \equiv  \langle\langle (\Delta \epsilon)^k \rangle\rangle = \frac{\partial^k \ln Z(\lambda)}{\partial (i\lambda)^k} \bigg{|} _{\lambda=0}.
\end{equation}
One can further define the energy current cumulants
\begin{equation}
\langle\langle (I^E)^k \rangle\rangle =\frac{\partial C_k(t)}{\partial t},
\end{equation}
which tend to the steady state energy current cumulants in the long time limit $t\rightarrow \infty$.
The second energy cumulant could be expressed as $C_2 (t) = \int_0^t dt_1 \int_0^t dt_2 \langle \delta I^E(t_1)\delta I^E(t_2) \rangle$, so that the second energy current cumulant is 
$\langle\langle (I^E)^2 \rangle\rangle = \frac{1}{2}\int_0^t dt_1 \langle \delta I^E(t_1)\delta I^E(t) \rangle + \frac{1}{2}\int_0^t dt_2 \langle \delta I^E(t) \delta I^E(t_2) \rangle$.
One should note that the second energy current cumulant $\langle\langle (I^E)^2 \rangle\rangle$ is not an average of a squared quantity. 
	To investigate statistical behaviors of the energy current through the left lead, we could focus on the energy operator which is actually the free Hamiltonian of the left lead $H_L$.
Under the two-time measurement scheme, GF of transferred energy in the left lead can be expressed over the Keldysh contour as \cite{RMP, JS3, gm2},
\begin{equation}\label{Z2}
Z(\lambda,t) =\mathrm{Tr}\left\{\rho(0)\mathcal{T}_C\exp\left[-\frac{i}{\hbar}\int_C H_\gamma(t')dt' \right]\right\}
=\mathrm{Tr}\left\{ \rho(0) U^\dag_{\lambda/2} (t,0) U_{-\lambda/2} (t,0) \right\},
\end{equation}
with the modified evolution operator ($\gamma = \pm \lambda/2$ depending on the branch of the contour, see Fig.~\ref{fig2}),
\begin{equation}\label{U}
  U_\gamma(t,0) = \mathcal{T} \exp\left[ -\frac{i}{\hbar}\int_{0}^{t} H_\gamma(t') dt'\right].
\end{equation}
Here the modified evolution operator is expressed by the modified Hamiltonian,
\begin{equation}  \label{mH}
H_\gamma = \bar{H}_{S} + \sum_k \Big[ \epsilon_{kL} c^\dag_{kL}(t_\gamma) c_{kL}(t_\gamma) + \epsilon_{kR} c^\dag_{kR} c_{kR} \Big] +\sum_k \Big[ \Big( t_{kL} c^\dag_{kL}(t_\gamma) X d + t_{kR} c^\dag_{kR} X d \Big)  + \mathrm{H.c.} \Big],
\end{equation}
with $t_\gamma=\hbar\gamma$, and $c_{kL}(t_\gamma) = e^{i\gamma H_L} c_{kL}(0) e^{-i\gamma H_L}$.

	GF for the transferred charges in transient regime has been expressed by NEGF in the time domain for the non-interacting case \cite{gm2} and in the polaronic regime using the DTA \cite{yeyati_FCS, BDong}.  GF for the energy current has expressed by NEGF and higher-order cumulants has been investigated by Yu \textit{et al.} for the non-interacting case \cite{gm3}. We now generalize the GF for the transferred energy to the interacting case in the polaronic regime following the derivation of the GF for transferred charges \cite{yeyati_FCS}. Following the procedure outlined in Ref. \onlinecite{Komnik}, one can get GF from the derivative of the logarithm of Eq.~\eqref{Z2} with respect to the counting field,
\begin{equation} \label{Z_der}
\frac{\partial\ln Z}{\partial \lambda} = \int_C dt' \sum_k \Big\langle {\cal T}_C \Big( t_{kL}
c_{kL}^\dag(t'\mp\hbar\lambda /2) X(t') d(t')
 - t_{kL}^* d^\dag(t') X^\dag (t') c_{kL}(t'\pm\hbar\lambda /2) \Big)\Big\rangle ,
\end{equation}
where we take $'-'$ in the first part and $'+'$ in the second for the forward time contour, while inversely for the backward contour (see Fig.~\ref{fig2}). The average $\big\langle {\cal T}_C
\cdots \big\rangle $ denotes $\mathrm{Tr}\left\{\rho(0)\mathcal{T}_C \cdots \exp\left[-\frac{i}{\hbar}\int_C H_\gamma(t')dt' \right]\right\}/Z(\lambda,t)$. The equation of motion of the three point Green function on the contour $\left\langle {\cal T}_C c_{kL}^\dag(t') X(t) d(t_2)\right\rangle$ is given by
\begin{equation}
\left(i\frac{\partial}{\partial t'} -\epsilon_{kL}\right)\left\langle {\cal T}_C c_{kL}^\dag(t') X(t) d(t_2) \right\rangle=t_{kL}^* \left\langle {\cal T}_C d^\dag(t') X^\dag(t') X(t) d(t_2) \right\rangle
\end{equation}
which could be written in an integral form \cite{Haug}
\begin{equation}
\left\langle {\cal T}_C c_{kL}^\dag(t') X(t) d(t_2) \right\rangle = \int_C dt_1
\left\langle {\cal T}_C d^\dag(t_1) X^\dag(t_1) X(t) d(t_2) \right\rangle t_{kL}^* g_{kL}(t_1,t')
\end{equation}

	Under DTA, one has the following decoupling \cite{yeyati_FCS}
\begin{equation}
\left\langle {\cal T}_C d^\dag(t_1) X^\dag(t_1) X(t) d(t_2) \right\rangle
\simeq \left\langle {\cal T}_C X^\dag(t_1) X(t) \right\rangle \left\langle {\cal T}_C d^\dag(t_1) d(t_2) \right\rangle = \Lambda(t,t_1) G(t_2,t_1) ,
\end{equation}
with $\Lambda(t,t_1) = \left\langle {\cal T}_C X^\dag(t_1) X(t) \right\rangle$ being the phonon cloud propagator which will be discussed later. Then we have
\begin{equation}
t_{kL}\left\langle {\cal T}_C c_{kL}^\dag(t') X(t') d(t') \right\rangle
= \int_C dt_1 G(t',t_1)\Lambda(t',t_1) \Sigma(t_1,t') .
\end{equation}
The self-energies due to the coupling to the leads under the DTA could be expressed as,
\begin{equation}
\Sigma_{\alpha,D}^{ab} (t_1,t_2) = \Sigma_{\alpha}^{ab} (t_1,t_2) \Lambda^{ba}_\alpha(t_2,t_1)
 = \Sigma_{\alpha}^{ab} (t_1,t_2) \Lambda^{ab}_\alpha(t_1,t_2),
\end{equation}
where $a,b=+,-$ denote different Keldysh components and
\begin{equation}
\Sigma_{\alpha}^{ab} (t_1,t_2)
= ab\theta(t_1)\theta(t_2) \sum_{k} t_{k\alpha}^* g^{ab}_{k\alpha}(t_1,t_2)  t_{k\alpha} .
\end{equation}
Note that the counting field enters the self-energy in absence of the phonon cloud operator and the modified self-energy can be expressed by \cite{gm3}
$\widetilde{\Sigma}^{ab}_L(t_1,t_2) = \Sigma^{ab}_L (t_1-t_2-(a-b)\hbar\lambda)$.
One can rewrite Eq.~\eqref{Z_der} as
\begin{equation}
\frac{\partial\ln Z}{\partial \lambda} = -\int_0^t dt_1 \int_0^t dt_2 {\rm Tr}_K \left\{
\frac{\partial \widetilde{\Sigma}_{L,D}(t_1,t_2) }{\partial\lambda} G(t_2,t_1) \right\},
\end{equation}
where ${\rm Tr}_K$ indicates the trace is over the Keldysh space.
Using the fact that $Z(\lambda=0,t)=1$, the GF could be expressed in the Fredholm determinant by the Keldysh NEGF in the time domain as \cite{yeyati_FCS, BDong_SC},
\begin{equation}  \label{gf}
Z(\lambda,t) = \det \left(G \widetilde{G}^{-1} \right)
\end{equation}
with
\begin{align}
G^{-1} = G_0^{-1} - \Sigma_{L,D}-\Sigma_{R,D} ,  \notag \\
\widetilde{G}^{-1} = G_0^{-1} -\widetilde{\Sigma}_{L,D}-\Sigma_{R,D} ,
\end{align}
where $G_0$ denotes the Green's function of the uncoupled QD, and the {\it tilde} indicates the inclusion of the counting field in the self-energy $\Sigma_{\alpha,D}$. Note that the Green's functions and self-energies without counting field possess the Keldysh structure,
\begin{equation}
A=\begin{pmatrix}
A^{++} & A^{+-} \\ A^{-+} & A^{--}
\end{pmatrix} .
\end{equation}
The phonon cloud operator $\Lambda^{ab}_\delta(t_1,t_2)$ that is coupled to lead $\delta=L,R$ is given by \cite{Mahan},
\begin{equation} \label{phonon}
\Lambda^{+-}_\delta(t_1,t_2) = \left[\Lambda^{-+}_\delta(t_1,t_2)\right]^*
=\sum_{m=-\infty}^{\infty} \alpha_{m\delta} e^{i m\omega_0 (t_1-t_2)},
\end{equation}
with
\begin{equation}
\alpha_{m\delta} = e^{-g^2(2n_{B\delta}+1)} e^{m\beta_\delta\omega_0/2} I_m\left( 2g^2\sqrt{n_{B\delta}(1+n_{B\delta})}\right),
\end{equation}
and $I_m$ being the modified Bessel function of the first kind, and Bose factor $n_{B\delta}=1/(e^{\beta_\delta\omega_0}-1)$, $\beta_\delta=1/k_B T_\delta$. We should mention that the temperature of the phonon cloud operator is dependent on which self-energy it multiplies with, and in the next section we will see that this will ensure the important fluctuation symmetry relation.
In the work by Y. Utsumi {\it et. al.}, a third thermal probe electrode due to the thermal bath was added to determine the temperature of the vibrations \cite{3rd-terminal}. In our work, we only consider the energy flow carried by electrons, and the fluctuation symmetry relation is already satisfied for the two-terminal system in Eq.~(43).  At zero-temperature $\alpha_m=\alpha_{mL}=\alpha_{mR}$ could be simplified as,
\begin{equation}
\alpha_m = 	\left \{ \begin{array}{cc}
e^{-g^2}g^{2m}/{m!}  \ \ \  & {\rm if} \ \ m\geq 0  \\  0  \ \ \  & {\rm if} \ \ m < 0
\end{array}
 \right. .
\end{equation}
The remaining components of $\Lambda_\delta$ could be calculated by the relations,
\begin{align}
&\Lambda^{++}_\delta(t_1,t_2)=\theta(t_1-t_2)\Lambda^{-+}_\delta(t_1,t_2)+\theta(t_2-t_1)\Lambda^{+-}_\delta(t_1,t_2), \notag \\
&\Lambda^{--}_\delta(t_1,t_2)=\theta(t_2-t_1)\Lambda^{-+}_\delta(t_1,t_2)+\theta(t_1-t_2)\Lambda^{+-}_\delta(t_1,t_2).
\end{align}
The Dyson equation bearing a Keldysh structure under DTA is
\begin{equation} \label{dyson1}
G = G_0 + G_0 \Sigma_{D} G ,
\end{equation}
where $ \Sigma_{D} = \Sigma_{L,D}+ \Sigma_{R,D}$.

	Utilizing the Dyson equation, Eq.~\eqref{gf} could be written as,
\begin{equation}
Z(\lambda,t) = \det \left[ I-G \left(\widetilde{\Sigma}_{L,D}-\Sigma_{L,D} \right) \right].
\end{equation}
so that CGF has the form,
\begin{equation} \label{transient_CGF}
\ln Z(\lambda,t) = {\rm Tr}\ln \left[ I-G \left(\widetilde{\Sigma}_{L,D}-\Sigma_{L,D} \right) \right],
\end{equation}
by using the relation $\det B=\exp[{\rm Tr}\ln B]$. Taking the first derivative of GF and noting that $\widetilde{\Sigma}_L^{+-}(t_1,t_2)=-\sum_k t_{kL}^* g_{kL}^{+-}(t_1-t_2-\lambda) t_{kL}$, energy current in the transient regime is found to be,
\begin{equation}
I_L^E(t) = \int_0^t dt' \left[ G^{+-}(t,t')\breve{\Sigma}^{-+}(t',t) - G^{-+}(t,t')\breve{\Sigma}^{+-}(t',t) \right] ,
\end{equation}
where
\begin{equation}
\breve{\Sigma}^{+-}(t',t)
=-\Lambda^{+-}(t'-t) \sum_k \epsilon_{kL} t_{kL}^* g_{kL}^{+-}(t'-t) t_{kL} ,
\end{equation}
and we have similar definition for $\breve{\Sigma}^{-+}(t',t)$.
The transient current expression formally agrees with the one which was obtained directly by NEGF method \cite{Yu}.

\section{Steady state energy transport FCS}
In the long-time limit, the system goes to steady state, and the Dyson equation Eq.~\eqref{dyson1} bearing the Keldysh structure in the energy domain could be expressed by
\begin{equation}
G = G_0 + G_0 \Sigma_{D} G ,
\end{equation}
so that \cite{BDong}
\begin{equation}
G = \frac{-1}{{\cal D}(\omega)} \left[ \begin{array}{cc}
-(\omega -\bar{\epsilon}) - \Sigma_{D}^{--}  &
\Sigma_{D}^{+-}  \\
\Sigma_{D}^{-+}  &
( \omega -\bar{\epsilon}) - \Sigma_{D}^{++}
\end{array}  \right] ,
\end{equation}
with
\begin{equation}
{\cal D}(\omega) = [\omega-\bar{\epsilon}-\Sigma_{D}^{r}(\omega)] [\omega-\bar{\epsilon}-\Sigma_{D}^{a}(\omega)] .
\end{equation}
The dressed retarded self-energy in frequency domain could be obtained by the Fourier transformation of the time domain counterpart with the form $\Sigma_D^r(t_1,t_2)=\theta(t_1-t_2)\left[\Sigma_D^{+-}(t_1,t_2)-\Sigma_D^{-+}(t_1,t_2)\right]$, so that in wide band limit (WBL) $W\rightarrow \infty$, \cite{BDong}
\begin{equation}  \label{dressed_retarded}
\Sigma_{\alpha, D}^{r}(\omega) =\sum_{m} \alpha_m\int \frac{dE}{2\pi} \frac{\Gamma_\alpha \left[1+f_{\alpha+m}(E)-f_{\alpha-m}(E)\right]}{\omega-E+i0^+} .
\end{equation}
The real and imaginary part could be obtained using Plemelj formula $1/(E\pm i0^+)=P(1/E)\mp i\pi\delta(E)$ which will be used in the numerical calculation. One can verify that the real part and imaginary part satisfies
\begin{align} \label{symmetry}
{\rm Im}\left[ \Sigma_{\alpha, D}^{r}(\mu_\alpha+\omega) \right] &= {\rm Im}\left[ \Sigma_{\alpha, D}^{r}(\mu_\alpha-\omega) \right]  , \notag \\
{\rm Re}\left[ \Sigma_{\alpha, D}^{r}(\mu_\alpha+\omega) \right] &= -{\rm Re}\left[ \Sigma_{\alpha, D}^{r}(\mu_\alpha-\omega) \right] ,
\end{align}
 respectively \cite{BDong}.

In the long-time limit, the Green's function and self-energy in Eq.~\eqref{transient_CGF} become time translation invariant so that scaled cumulant generating function (SCGF) ${\cal F}(\lambda)=\lim_{t\rightarrow \infty} \ln Z(\lambda)/t$ could be expressed in the energy domain as
\begin{equation} \label{SCGF}
{\cal F}(\lambda)=\int \frac{d\omega}{2\pi}\ln\bigg\{ 1+\sum_{mn} T_{mn}(\omega)[f_{L+m}(1-f_{R-n})
(e^{i\lambda\omega}-1)+f_{R+n}(1-f_{L-m}) (e^{-i\lambda\omega}-1) ]\bigg\}.
\end{equation}
In this expression $T_{mn}(\omega)$ is the transmission coefficient involving $m$ and $n$ vibrational quanta in the left and right lead, respectively, with the form,
\begin{equation}
T_{mn}(\omega) = \frac{\Gamma_L\Gamma_R\alpha_m\alpha_n}{{\cal D}(\omega)} .
\end{equation}
Taking the first order derivative of SCGF with respect to $\lambda$, we can get the expression of energy current,
\begin{equation} \label{IE}
\langle I^E \rangle =\int\frac{d\omega}{2\pi} \hbar\omega \sum_{mn} T_{mn}(\omega)[f_{L+m}(1-f_{R-n})  -f_{R+n}(1-f_{L-m})].
\end{equation}

Now we consider the universal relations for energy current cumulants under finite temperature gradient with zero bias which is in analogy with the universal relation for particle current cumulants \cite{Tobiska, Forster}. Using the relation $\alpha_{-m}=e^{-\beta_L m\omega_0}\alpha_m$,  $\alpha_{-n}=e^{-\beta_R n\omega_0}\alpha_n$ and $f_R(1-f_L)=\exp(\Delta\beta\omega) f_L(1-f_R)$ with $\Delta\beta =\beta_L-\beta_R$ for $\Delta\mu=0$ in Eq.~\eqref{SCGF}, we have the fluctuation symmetry relation
\begin{equation}
{\cal F}(\xi) = {\cal F}(-\xi + \Delta\beta ) \label{47}
\end{equation}
with $i\lambda$ being replaced by $\xi$ for convenience. One can verify that the fluctuation symmetry can only be satisfied by considering the dependency of phonon temperature with respect to the specific lead. 
In the linear response regime $\Delta\beta \rightarrow 0$, we can expand both sides as Taylor series around $\Delta\beta =0$ and $\xi = 0$, which leads to,
\begin{equation} \label{Taylor}
\frac{d^k {\cal F}(-\xi+\Delta\beta , \Delta\beta)}{d \Delta\beta^k} \bigg|_0
=\sum_{l=0}^k \binom{k}{l} \frac{\partial^k {\cal F}(\xi , \Delta\beta)}{\partial \Delta\beta^{k-l} \partial \xi^l} \bigg|_0 . 
\end{equation}
where we have written the dependence of $\Delta\beta$ of SCGF explicitly out in both sides.
Since ${\cal F}(\xi=0,\Delta\beta) = 0$, Eq.(\ref{47}) gives ${\cal F}(\Delta\beta,\Delta\beta) = 0$, from which we find that the LHS of Eq.(\ref{Taylor}) vanishes. 
The last term in the summation of Eq.~\eqref{Taylor} is the $k$th derivative of the SCGF with respect to the counting field $\xi$, which is actually $\langle\langle (I^E)^k \rangle\rangle$ at equilibrium. Then we have the relation
\begin{equation}
\langle\langle (I^E)^k \rangle\rangle_{\rm eq} = - \sum_{l=1}^{k-1} \binom{k}{l}
\frac{\partial^{k-l} \langle\langle (I^E)^l \rangle\rangle}{\partial \Delta\beta^{k-l}} ,
\end{equation}
in which the energy current cumulant at equilibrium is expressed by a linear combination of lower order energy current cumulants. This is similar to the case that the particle current cumulant could could be expressed by a linear combination of lower order particle current cumulants in the presence of small voltage bias \cite{Tobiska, Forster}.

	We now show numerical calculations regarding steady state energy current cumulants under temperature gradient and external bias of molecular junctions in the polaronic regime. The energies are measured in the unit of vibron frequency $\omega_0$, and the linewidth amplitude is chosen to be $\Gamma = 0.05\omega_0$ which indicates weak coupling. In addition, WBL is taken in our steady state calculation.

\begin{figure}
  \includegraphics[width=4.0in]{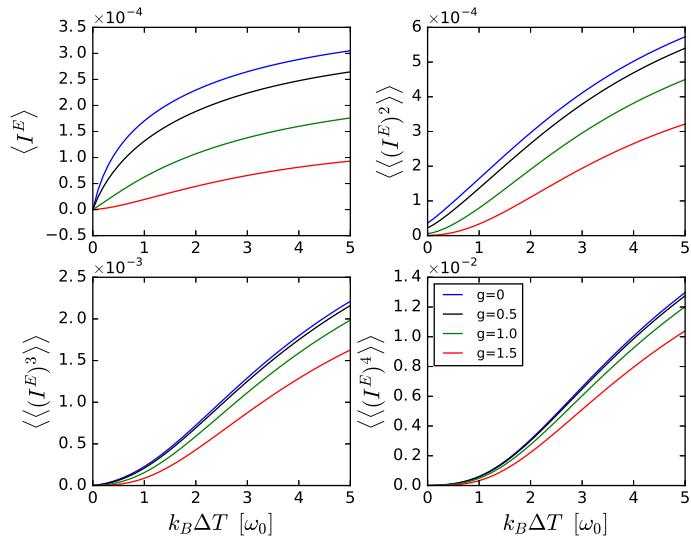}\\
  \caption{1st to 4th energy current cumulants for increasing $g$ (0 (blue), 0.5 (green), 1.0 (black) and 1.5 (red)) versus temperature gradient $\Delta T = T_L - T_R$ with the left lead warmer and temperature of right lead fixed at $k_B T_R=0.2\omega_0$. The renormalized energy level of the QD is $\bar{\epsilon}=0$.}
  \label{fig3}
\end{figure}

	The first to fourth energy current cumulants for increasing $g$ versus temperature gradient $\Delta T = T_L - T_R$ with the left lead warmer and temperature of right lead fixed at $k_B T_R=0.2\omega_0$ are shown in Fig.~(\ref{fig3}). The chemical potentials in both leads are set to be zero and the renormalized energy level of the QD is $\bar{\epsilon}=0$. The energy current cumulants become smaller with the increasing of $g$ because of the suppression of transport due to electron-phonon interaction.  The second energy current cumulant with zero temperature gradient is finite due to the thermal noise in the leads, and it is reduced with increasing $g$.
	
\begin{figure}
  \includegraphics[width=4.0in]{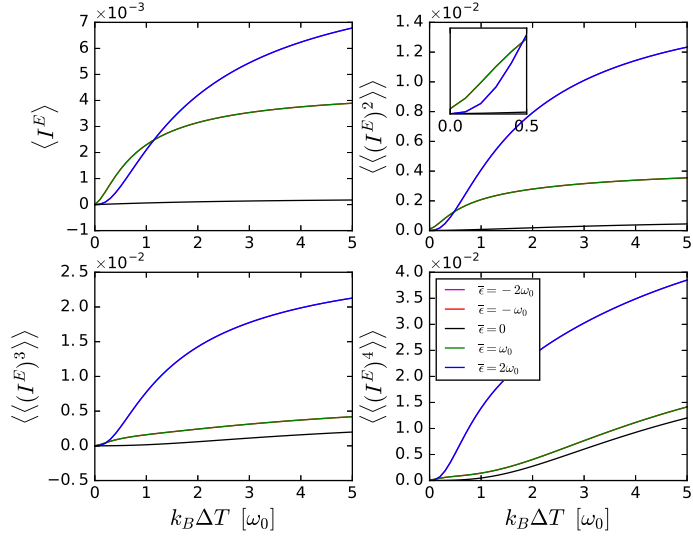}\\
  \caption{1st to 4th energy current cumulants for different renormalized energy levels of the QD $\bar{\epsilon}$ ($-2\omega_0$ (magenta), $-\omega_0$ (red), $0$ (black) $\omega_0$ (green) and $2\omega_0$ (blue)) versus temperature gradient $\Delta T = T_L - T_R$ with $k_B T_R=0.2\omega_0$. $g=1.0$. 1st to 4th cumulants are even functions of $\bar{\epsilon}$. The lines of $\bar{\epsilon}=-2\omega_0$ coincide with the lines of  $\bar{\epsilon}=2\omega_0$, and the lines of $\bar{\epsilon}=-\omega_0$ coincide with the lines of  $\bar{\epsilon}=\omega_0$. }
  \label{fig4}
\end{figure}

In Fig.~(\ref{fig4}), energy current cumulants with different renormalized energy levels of the QD with $g=1$ are plotted. We can see that the first to fourth cumulants and SCGF as well are even functions of $\bar{\epsilon}$. This can be understood as follows.
Since the chemical potentials of both leads are zero, one can set
\begin{equation}
X_{mn}(\omega) = f_{L+m}(\omega)[1-f_{R-n}(\omega)] e^{i\lambda\omega} + f_{R+n}(\omega) [1-f_{L-m}(\omega)] e^{-i\lambda\omega} ,
\end{equation}
and verify that,
\begin{equation}
X_{mn}(\omega) =  X_{mn}(-\omega) ,
\end{equation}
using the relation $f_{L+m}(\omega) = 1-f_{L-m}(-\omega)$.
In the WBL, from Eq.~\eqref{symmetry}, the real and imaginary part of the dressed retarded self-energy are the odd and even function of $\omega$, respectively, so that we have the following symmetry with respect to the transmission coefficient in the polaronic regime
\begin{equation}  \label{symmetry_T}
T_{mn}(\omega,\bar{\epsilon}) = T_{mn}(-\omega,-\bar{\epsilon})  .
\end{equation}
where the dependency of $\bar{\epsilon}$ has been written explicitly.
Then, we have the following symmetry of SCGF with respect to $\bar{\epsilon}$,
\begin{equation}
{\cal F}(\lambda ,\bar{\epsilon}) = {\cal F}(\lambda ,-\bar{\epsilon}) .
\end{equation}
with $\mu_L=\mu_R=0$ in the WBL. One can also see from Fig.~(\ref{fig4}), $\langle I^E\rangle (\bar{\epsilon}=2\omega_0)$ is smaller than $\langle I^E\rangle (\bar{\epsilon}=\omega_0)$ under small temperature gradient, and this is also for the second energy current cumulant.  Since the linewidth amplitude $\Gamma=0.05\omega_0$ is small, so that the transmission coefficient which is centered around $\bar{\epsilon}$ is narrow. As a result, the main contribution to the transport process is coming from energy near $\bar{\epsilon}$. When the temperature gradient across the junction is small, the difference of Fermi distribution functions between left and right lead $f_L(\omega)-f_R(\omega)$ is smaller near $\bar{\epsilon}=2\omega_0$ than near $\bar{\epsilon}=\omega_0$.  When $T_L$ increases, the difference of Fermi distribution functions between left and right lead $f_L(\omega)-f_R(\omega)$ near $\omega=2\omega_0$ could exceed the difference near $\omega=\omega_0$, so that the first and second cumulant with larger $\bar{\epsilon}$ is larger than the ones with smaller $\bar{\epsilon}$.

\begin{figure}
  \includegraphics[width=4.0in]{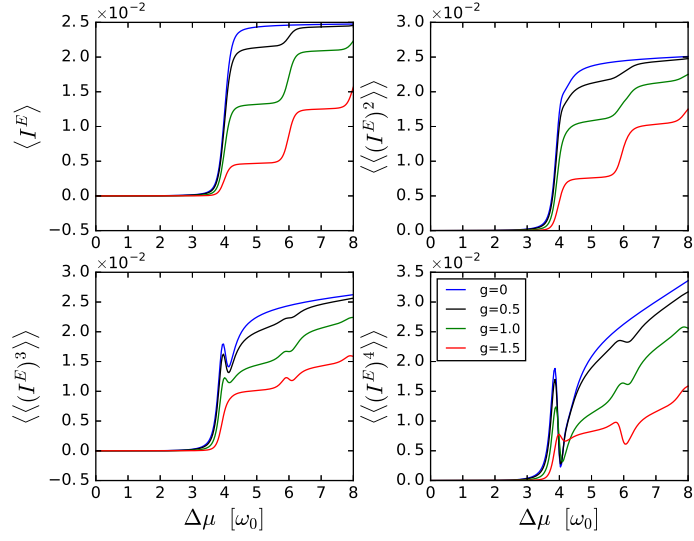}\\
  \caption{1st to 4th energy current cumulants for increasing $g$ (0 (blue), 0.5 (green), 1.0 (black) and 1.5 (red)) versus external bias $\Delta\mu$ with $\mu_L=\Delta\mu/2$ and $\mu_R=-\Delta\mu/2$. Temperatures of both leads are $k_B T_L=k_B T_R=0.04\omega_0$. The renormalized energy level of the QD is chosen to be $\bar{\epsilon}=2\omega_0$.}
  \label{fig5}
\end{figure}

The first to fourth energy current cumulants for increasing $g$ versus external bias $\Delta\mu$ with $\mu_L=\Delta\mu/2$ and $\mu_R=-\Delta\mu/2$ are shown in Fig.~(\ref{fig5}). Temperatures of both leads are chosen to be very small with $k_B T_L=k_B T_R=0.04\omega_0$ which is almost in the regime of zero temperature. The renormalized energy level of the QD is $\bar{\epsilon}=2\omega_0$. For the non-interacting case, the energy current and second cumulant are almost zero when bias is below $\Delta\mu =4\omega_0=2\bar{\epsilon}$ and display plateau structures when the external bias exceeds $2\bar{\epsilon}$.
The width of transmission coefficient is small due to the small linewidth amplitude $\Gamma = 0.05\omega_0$. When $\Delta\mu =2\bar{\epsilon}$, chemical potential of the left lead is equal to the renormalized energy of QD, $\mu_L = \bar{\epsilon}$, in which energy the transmission coefficient experiences a sharp increase and reaches its largest value as indicated in Fig.~\ref{fig1}(b).
From the Fig.~\ref{fig5}, we observe electron-phonon coupling enables the plateau height to become smaller, however creates smaller steps at $\Delta\mu =2\bar{\epsilon}+2n\omega_0$ with $n=1,2,3 \cdots$. This is due to the presence of sidebands in the leads and could be understood as follows. In the presence of the polaronic regime, from Eq.~\eqref{IE}, we can approximately write the energy current in presence of bias voltage at zero temperature as, (ignore the terms with product of Fermi distribution function)
\begin{equation} \label{step}
\langle I^E \rangle \approx\int\frac{d\omega}{2\pi} \hbar\omega \sum_{m\geq 0} T_{m}(\omega)(f_{L+m} -f_{R+m}) = \int_{-\Delta\mu/2}^{\Delta\mu/2}\frac{d\omega}{2\pi} \hbar\omega T_{0} +
\int_{-\Delta\mu/2 -\omega_0}^{\Delta\mu/2-\omega_0}\frac{d\omega}{2\pi} \hbar\omega T_{1} +
\int_{-\Delta\mu/2 -2\omega_0}^{\Delta\mu/2-2\omega_0}\frac{d\omega}{2\pi} \hbar\omega T_{2}+\cdots ,
\end{equation}
with $T_m = \frac{\Gamma_L\Gamma_R\alpha_m}{{\cal D}(\omega)} \propto \alpha_m = e^{-g^2} g^{2m}/m!$. The energy current is written as a sum of a series, with each term coming from a different sideband in the leads. The first plateau of the energy current in the polaronic regime is mainly due to the first term in Eq.~\eqref{step}, and the second plateau due to the contribution from the second term in Eq.~\eqref{step} with one polaron involved in the transport process, and etc.. We find that $T_m/T_{m-1}= g^2/m$ is responsible for the ratios between plateau heights. One can see that when $g=0.5$, $T_1/T_0 = 0.25$, so that the height of the second plateau is a quarter of that of the first plateau at zero temperature, which explained what we see in Fig.~\ref{fig5}. This is also applicable to the case $g=1.0$ with $T_1/T_0 = 1.0$ and the case $g=1.5$ with $T_1/T_0 = 2.25$. One should note that the temperature of the system in Fig.~\ref{fig5} is very small.

The plateau structures disappear in the third and fourth energy current cumulants. Instead a dip occurs at $\Delta\mu =2\bar{\epsilon}$ for both the third and fourth energy current cumulants with fourth cumulant larger for both non-interacting and interacting cases. Polaronic regime creates smaller dips at $\Delta\mu =2\bar{\epsilon}+2n\omega_0$ with $n=1,2,3 \cdots$ which could also be identified in Fig.~\ref{fig7}. Increasing $g$ reduces the amplitude of the dip at $\Delta\mu =2\bar{\epsilon}$ but increases the amplitude at $\Delta\mu =2\bar{\epsilon}+2n\omega_0$. The explanation is as follows. For the non-interacting case under zero-temperature, we have
\begin{align}
\langle I^E \rangle &= \int \frac{d\omega}{2\pi} \omega T(\omega) ,
\notag \\
\langle\langle (I^E)^2 \rangle\rangle &= \int \frac{d\omega}{2\pi} \omega^2 T(\omega)[1-T(\omega)] ,  \\
\langle\langle (I^E)^3 \rangle\rangle &= \int \frac{d\omega}{2\pi} \omega^3 T(\omega)[1-T(\omega)][1-2T(\omega)] ,\notag \\
\langle\langle (I^E)^4 \rangle\rangle &= \int \frac{d\omega}{2\pi} \omega^4 T(\omega)[1-T(\omega)][1-6T(\omega)+6T^2(\omega)] \notag  ,
\end{align}
with the ranges of integration from $-\Delta\mu/2$ to $\Delta\mu/2$.
We can further take derivative of $\langle\langle (I^E)^k \rangle\rangle$ with respect to external bias $\Delta\mu$, ${\partial \langle I^E \rangle}/{\partial \Delta \mu}$ and ${\partial \langle\langle (I^E)^2 \rangle\rangle}/{\partial \Delta \mu} $ is always positive definite since the transmission coefficient for non-interacting case has the form $T(\omega)=\frac{\Gamma^2/4}{(\omega-\bar{\epsilon})^2+\Gamma^2/4}$. However the derivative of the third and fourth cumulant with respect to external bias change sign around $\Delta\mu =2\bar{\epsilon}$ and also the transmission coefficient experiences an abrupt change because of small linewidth amplitude. This leads to the the dips of third and forth cumulant of energy current as shown in Fig.~\ref{fig5}.

\begin{figure}
  \includegraphics[width=4.0in]{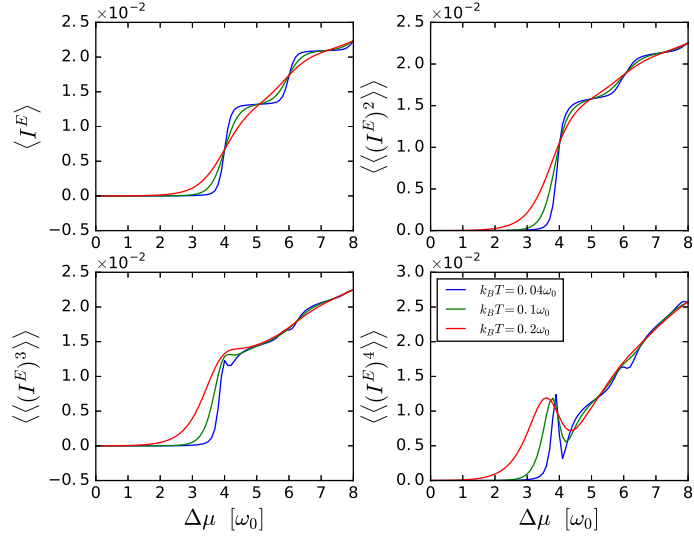}\\
  \caption{1st to 4th energy current cumulants for increasing temperatures $k_BT$ ($0.04\omega_0$ (blue), $0.1\omega_0$ (green) and $0.2\omega_0$ (red)) versus external bias $\Delta\mu$ with $\mu_{L(R)}=\pm \Delta\mu/2$. $g=1.0$ and $\bar{\epsilon}=2\omega_0$.}
  \label{fig6}
\end{figure}

The influence of temperature on cumulants under external bias is depicted in Fig.~\ref{fig6}, one can see that both the plateaus and dips get smoothed or even disappeared when temperature increases.
In Fig.~\ref{fig7}, energy current cumulants with different $\bar{\epsilon}$ with $g=1$ are plotted. We can see that the first and third cumulants are odd functions of $\bar{\epsilon}$, while the second and fourth cumulants are even functions of $\bar{\epsilon}$. The reason is as follows. Under zero temperature, the transport is unidirectional and Fermi-Dirac distribution function $f_{L(R)}$ has a step-wise form, since the transmission coefficient is peaked around the resonant level $\bar{\epsilon}$ with a very small linewidth amplitude (say $\delta \epsilon$), the energies of electron which mainly contribute to the energy transport are very close to $\bar{\epsilon}$. So if we change the sign of $\bar{\epsilon}$ from positive to negative, then most of electron energies will reverse their signs if $\delta \epsilon <{\bar \epsilon}$. Since the energy current is proportional to energies of electron, this will lead to the energy current reversal.

\begin{figure}
  \includegraphics[width=4.0in]{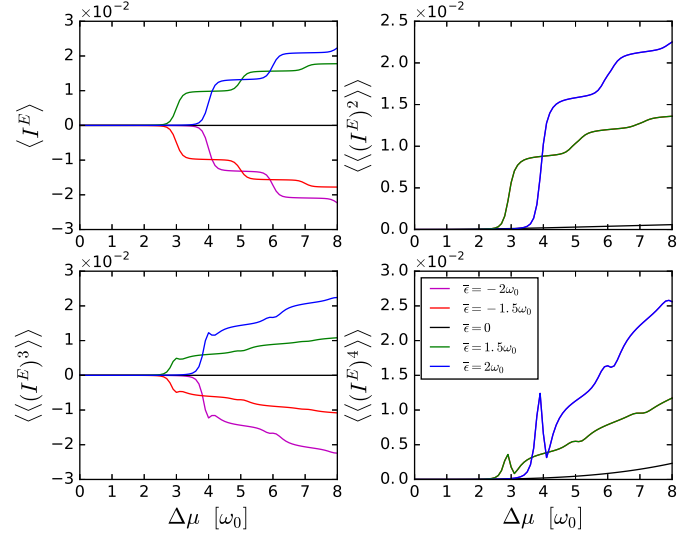}\\
  \caption{1st to 4th energy current cumulants for different $\bar{\epsilon}$ ($-2\omega_0$ (magenta), $-1.5\omega_0$ (red), $0$ (black) $1.5\omega_0$ (green) and $2\omega_0$ (blue)) versus external bias $\Delta\mu$ with $\mu_L=\Delta\mu/2$ and $\mu_R=-\Delta\mu/2$. $g=1.0$ and temperatures of both leads are $k_B T_L=k_B T_R=0.04\omega_0$. The lines of 2nd and 4th cumulant of $\bar{\epsilon}=-2\omega_0$ coincide with the lines of $\bar{\epsilon}=2\omega_0$, and the lines of 2nd and 4th cumulant of $\bar{\epsilon}=-\omega_0$ coincide with the lines of $\bar{\epsilon}=\omega_0$.}
  \label{fig7}
\end{figure}

\section{Transient Dynamics of energy transport}
We first investigate the behaviors of energy current at very short time. To do that, we expand the GF to the lowest order in time,
\begin{widetext}
\begin{equation}
Z(\lambda ,t)\approx 1+ \int_0^t dt_1 \int_0^t dt_2
\Big[\widetilde{\Sigma}_{L,D}^{-+}(t_1,t_2) - \Sigma_{L,D}^{-+}(t_1,t_2)\Big] G_0^{+-}(t_2,t_1)
+\Big[\widetilde{\Sigma}_{L,D}^{+-}(t_1,t_2) -\Sigma_{L,D}^{+-}(t_1,t_2)\Big] G_0^{-+}(t_2,t_1)
\end{equation}
\end{widetext}
The expressions of Green's function for isolated QD and self-energy are given in the Appendix. Under the wide-band limit $W\rightarrow \infty$, we can obtain the GF in the short time limit in a compact form as,
\begin{equation} \label{shortZ}
Z(\lambda,t)\approx 1+ A_{L0}(n_d-1) + A_{L1}n_d ,
\end{equation}
where
\begin{align}
&A_{L0} = \frac{\Gamma_L}{\pi}\sum_{n=-\infty}^{\infty}\alpha_n \int d\omega
 (e^{i\omega\lambda}-1) M(\omega) f_{L+n}(\omega) ,
\notag \\
&A_{L1} = \frac{\Gamma_L}{\pi}\sum_{n=-\infty}^{\infty}\alpha_n \int d\omega
  (e^{-i\omega\lambda}-1) M(\omega) [f_{L-n}(\omega)-1] ,
\end{align}
with
\begin{equation}
M(\omega) = \frac{1-\cos[(\omega-\bar{\epsilon})t]}{(\omega-\bar{\epsilon})^2}.
\end{equation}
From now on we use $f_{\alpha\pm n}$ to denote $f_\alpha (\omega\pm n\omega_0)$. We can see from the expression of short time limit of the GF that the transport process is unidirectional in the short time limit. We can get the current expressions in the short time limit as 
\begin{equation}
I_L^E(t)= \frac{d}{dt}\frac{\partial \ln Z(\lambda,t)}{\partial(i\lambda)}\Big|_{\lambda=0} =
\frac{\Gamma_L}{\pi}\sum_{n=-\infty}^{\infty}\alpha_n \int d\omega \frac{\omega\sin[(\omega-\bar{\epsilon})t]}{\omega-\bar{\epsilon}} \Big\{ f_{L+n}(\omega) (n_d-1) - [f_{L-n}(\omega)-1] n_d \Big\} .
\end{equation}

	We apply the formalism to perform numerical calculation with respect to the transient dynamics of energy current under temperature gradient and external bias, respectively. The energies are measured in the unit of $\Gamma$ and $1/\Gamma$ is the unit of time. We only consider the case where the QD is initially unoccupied $n_d=0$ and the linewidth amplitude in Eq.~\eqref{linewidth} is set to be $\Gamma_L = \Gamma_R = \Gamma/2$ and the bandwidth is also set to be the same for both leads with $W=10\Gamma$.
	
\begin{figure}
  \includegraphics[width=4.0in]{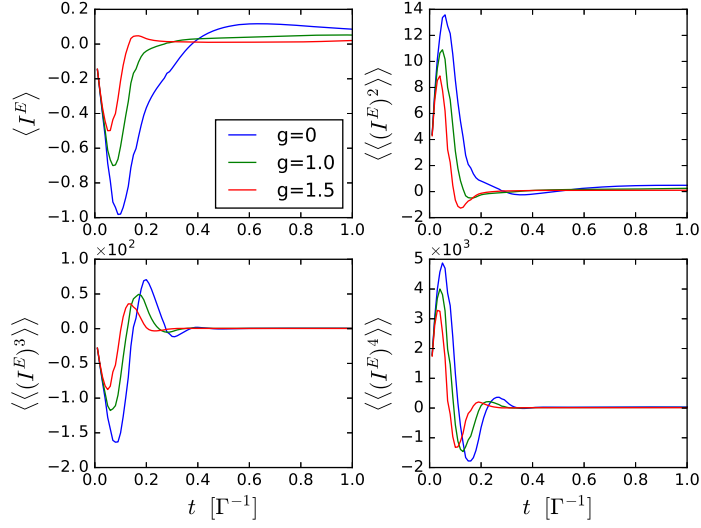} \\
  \caption{1st to 4th transient energy current cumulants in the left lead for increasing $g$ (0 (blue), 1.0 (green) and 1.5 (red)) for initially empty QD under temperature gradient between the leads. The energies are measured in the unit of $\Gamma$ and $1/\Gamma$ is the unit of time. The temperatures of the two leads are chosen to be $k_B T_L=1.5\Gamma$ and $k_B T_R=1.2\Gamma$ with the left lead warmer. The renormalized energy level of the QD is $\bar{\epsilon}=2\Gamma$ and the frequency of the localized vibron $\omega_0=6\Gamma$. }
  \label{fig8}
\end{figure}

\begin{figure}
  \includegraphics[width=4.0in]{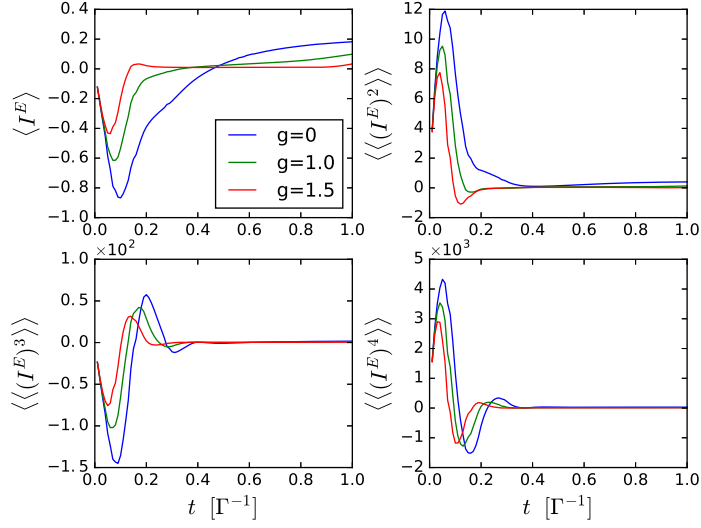}\\
  \caption{1st to 4th transient energy current cumulants in the left lead for increasing $g$ (0 (blue), 1.0 (green) and 1.5 (red)) for initially empty QD under external bias at zero temperature. The energies are measured in the unit of $\Gamma$ and $1/\Gamma$ is the unit of time. The chemical potential of the left and right lead are chosen to be $\mu_{L(R)}=\pm 2\Gamma$. The renormalized energy level of the QD is $\bar{\epsilon}=1.5\Gamma$ and the frequency of the localized vibron $\omega_0=6\Gamma$. }
  \label{fig9}
\end{figure}

	First to fourth transient energy current cumulants, $\langle\langle (I^E)^k\rangle\rangle$ for $k=1,2,3,4$, in the left lead for increasing $g$ under temperature gradient and external bias are shown, respectively, in Fig.~\ref{fig8} and in Fig.~\ref{fig9}. Increasing $g$ corresponds to the increasing of the electron-phonon coupling strength. The frequency of the localized vibron is $\omega_0=6\Gamma$. The renormalized energy level of the QD is $\bar{\epsilon}=2\Gamma$ for case under temperature gradient and $\bar{\epsilon}=1.5\Gamma$ for the case with external bias. The left lead is assumed to be warmer with the temperatures of the two leads to be $k_B T_L=1.5\Gamma$ and $k_B T_R=1.2\Gamma$ while the chemical potentials in both leads are set to zero in the case under temperature gradient. The temperature parameter in the phonon cloud operator Eq.~\eqref{phonon} should be the temperature of the lead where the phonon cloud operator acts. 
For the case under external bias, the chemical potential of the left and right lead are chosen to be $\mu_L=2\Gamma$ and $\mu_R=-2\Gamma$. The temperature of both leads is zero, while a small temperature $k_B T=0.1\Gamma$ in the phonon cloud operator is taken in order to stabilize the numerical calculations. 	
	
\begin{figure}
  \includegraphics[width=4.0in]{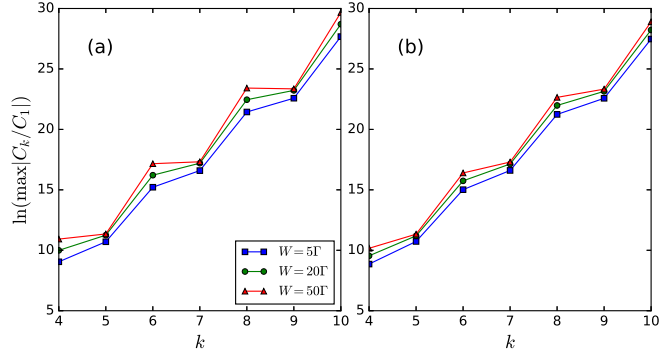}\\
  \caption{Logarithm of maximum amplitude of the normalized transient energy cumulants $M_k = {\rm max} |C_k/C_1|$ at short times versus $k$ for different bandwidths $W$ under (a) temperature gradient, and (b) external bias. Maximum amplitudes $M_k$ for different interaction parameters $g=0, 1.0$ and $1.5$ coincide.}
  \label{fig10}
\end{figure}

	As a general feature for both the non-interacting ($g=0$) and interacting cases, the transient amplitudes of $\langle\langle (I^E)^k\rangle\rangle$ increase with cumulants order. This behavior is universal and will be investigated in detail in Fig.~\ref{fig10}. The second and fourth energy current cumulants may even oscillate to negative values at short times. 
	The negativity of the second energy current cumulants can be explained as follows. The energy cumulant $C_2(t)$ must be positive at all times from a statistical view, however it can oscillate at short times so that the second energy current cumulants which is the derivative of $C_2(t)$ may not be positive at short times. $\langle\langle (I^E)^2 \rangle\rangle$ at steady state (long time limit) is positive and can be identified from the figures.	
	 The amplitudes of oscillation in the evolution and the asymptotic values of the cumulants are suppressed with the increasing of $g$. The first and third energy current cumulants in the stationary limit are positive under temperature and external bias, since we put the normalized energy level of QD above the Fermi energy of the both leads so that the electrons with positive energy contribute to the transport process. However, in short times, the energy current and third cumulant oscillate to negative values with a minimum. This could be understood as follows. Since the QD is prepared initially empty, once the system is connected, the contribution to the transport process mainly comes from electron of the left lead which could be seen from Eq.~\eqref{shortZ}. The contribution of energy current cumulants from the energy window $[0,\mu_L]$ cancels with the contribution from $[-\mu_L,0]$, so that energy below $-\mu_L$ in the left lead will contribute to the energy transport process which leads to the negativity of the first and third energy current cumulants in the short times. The cumulants of transient energy current approach to their steady state values in the long time limit.

	We also plot the logarithm of maximum amplitude of the normalized transient energy cumulants $M_k = {\rm max} |C_k/C_1|$ under temperature gradient [Fig.~\ref{fig10}(a)] and external bias [Fig.~\ref{fig10}(b)]. Different lines with respect to different bandwidths $W$ are plotted, while the other parameters are same as in Fig.~\ref{fig8} and Fig.~\ref{fig9}. 
Maximum amplitudes $M_k$ for different interaction parameter $g=0, 1.0$ and $1.5$ coincide. 
We can see from the figure that both $\ln(M_{2k})$ and $\ln(M_{2k+1})$ are linear with cumulants order $k$ with the slope close to $3$ but they have different intercepts. This universal scaling of normalized transient energy cumulants is found under both the temperature gradient and external bias, and it is the result of the universality of the GF in the short time which was also reported in the charge cumulants \cite{yeyati_FCS, exp1}. Theoretical understanding of this behavior for the noninteracting case was reported in our previous work \cite{gm3}. Interestingly, turning on the electron-phonon interaction does not affect this behavior.

\section{Conclusion}\label{sec4}
Both steady state and transient behaviors of energy transport carried by electrons in molecular junctions for the Anderson-Holstein model in the polaronic regime have been investigated using FCS. Using two-time measurement scheme and equation of motion technique, GF for the energy current could be expressed as a Fredholm determinant in the time domain using NEGF. The DTA decoupling scheme \cite{DTA} which could provide a good description in dealing with the phonon cloud operator has been adapted in obtaining GF. This formalism allows us to analyze the time evolution of energy transport dynamics after a sudden switch of the coupling between the dot and the leads towards the stationary state.
The amplitudes of oscillation in the evolution and the asymptotic values of the cumulants are suppressed with the increasing of $g$. The universal scaling of normalized transient energy cumulants is found under external bias.

	In the steady states, universal relations for energy current cumulants under finite temperature gradient with zero bias and this enables us to express the equilibrium energy current cumulant by a linear combination of lower order cumulants. Behaviors of energy current cumulants (from the first to the fourth) under temperature gradient and external bias are numerically shown and explained.
	Under external bias, the energy current and second cumulant are almost zero when bias is below $\Delta\mu =2\bar{\epsilon}$ for the non-interacting case and display plateau structures when the external bias exceeds $2\bar{\epsilon}$.
Due to the sidebands in leads in polaronic regime, the plateau heights become smaller, however smaller plateau steps appear at $\Delta\mu =2\bar{\epsilon}+2n\omega_0$ with $n=1,2,3 \cdots$.
The plateau structures disappear in the third and fourth energy current cumulants. Instead a dip occurs at $\Delta\mu =2\bar{\epsilon}$ for both the third and fourth energy current cumulants with fourth cumulant larger for both non-interacting and interacting cases. Polaronic regime creates smaller dips at $\Delta\mu =2\bar{\epsilon}+2n\omega_0$ with $n=1,2,3 \cdots$.

\begin{acknowledgments}
This work was financially supported by NSF-China under Grant No. 11374246, the General Research Fund (Grant No. 17311116), and the University Grant Council (Contract No. AoE/P-04/08) of the Government of HKSAR.
\end{acknowledgments}

\appendix
\section*{APPENDIX: Green's function and self-energy in the time domain}
Description on how to calculate the uncoupled dot Green's function and the self-energy in the time domain in the absence of the phonon cloud operator is sketched here.
The four correlation functions of the uncoupled dot are given in the book by A. Kamenev, \cite{Ka2}
\begin{align}
 i G_0^{+-}(t_1,t_2) &= -n_d\exp\{-i\bar{\epsilon} (t_1-t_2)\} \notag  \\
 i G_0^{-+}(t_1,t_2) &= (1-n_d)\exp\{-i\bar{\epsilon} (t_1-t_2)\} \notag  \\
 i G_0^{++}(t_1,t_2) &= \theta(t_1-t_2)iG_0^{-+}+\theta(t_2-t_1)iG_0^{+-} \notag  \\
 i G_0^{--}(t_1,t_2) &= \theta(t_2-t_1)iG_0^{-+} +\theta(t_1-t_2)iG_0^{+-} ,
\end{align}
where $n_d$ is the initial occupation number of the QD before the system is connected.
Lorentzian linewidth function with the linewidth amplitude $\Gamma_\alpha$ and band width $W$,
\begin{equation}
{\bf \Gamma}_{\alpha}(\omega)=\frac{\Gamma_{\alpha} W^2}{\omega^2+W^2},
\end{equation}
is used to describe the self-energy $\Sigma_{L(R)}$ in absence of the phonon cloud operator, so that the numerical calculation would be more realistic.
The equilibrium energy dependent self-energy can be written as,
\begin{equation}  \label{selfenergy}
{\Sigma}^r_\alpha(\omega) = \frac{\Gamma_\alpha W}{2(\omega + iW)} .
\end{equation}
Performing Fourier transform, the retarded self-energy in the time domain could be obtained,\cite{gm2}
\begin{equation}  \label{sigmar}
\Sigma^r_{\alpha} (t_1, t_2) = -\frac{i}{2} \theta(t_1-t_2) \Gamma_\alpha W e^{-(i\mu_\alpha + W)(t_1 - t_2)},
\end{equation}
where $\mu_\alpha$ is the chemical potential of the $\alpha$-lead.
	For the lesser self-energy in the time domain,
\begin{equation}
\Sigma_{\alpha}^{<}(t_1,t_2)=i\int\frac{d\omega}{2\pi} e^{-i\omega(t_1-t_2)}
f_\alpha(\omega){\bf \Gamma}_L(\omega-\mu_\alpha)
\end{equation}
with $f_{\alpha}(\omega)=1/\left[e^{\beta(\omega-\mu_\alpha)}+1\right]$. It is a function of the time difference, and one can let $\tau=t_1-t_2$ for convenience. When $t_1=t_2$,
\begin{equation}
\Sigma_{\alpha}^{<}(t_1,t_2)=\frac{i}{4}\Gamma_{\alpha} W   .
\end{equation}
The case of $t_1>t_2$ for both the zero and non-zero temperature is to be considered first. At non-zero temperature, if $t_1>t_2$, it has poles $\frac{-i(2n+1)\pi}{\beta_{\alpha}}$ and $-iW$, where $n=0,1,2,3...$, so that,
\begin{widetext}
\begin{align}
&\Sigma_{\alpha}^{<}(t_1,t_2) = \frac{i\Gamma_{\alpha} W}{2} e^{-i\mu_{\alpha} \tau}
 \left\{ e^{-W\tau} \left[1+\frac{E1(-W\tau)}{2i\pi} \right]
 - e^{W\tau} \frac{E1(W\tau)}{2i\pi} \right\}  \qquad k_BT_\alpha=0 ,  \notag \\
&\Sigma_{\alpha}^{<}(t_1,t_2)= \frac{i\Gamma_{\alpha} W}{2} e^{-i\mu_{\alpha} \tau} \left\{
\frac{\exp(-W\tau)}{\exp(-i\beta_{\alpha} W)+1} -\frac{2}{i\beta_{\alpha}}
\sum_{n=0}^{+\infty}\exp\left[-\frac{(2n+1)\pi}{\beta_{\alpha}}\tau \right]\frac{W}{W^2-\left[\frac{(2n+1)\pi}{\beta_{\alpha}}\right]^2} \right\} \qquad k_BT_\alpha\neq 0 ,
\end{align}
\end{widetext}
where $E1(x)=\int_x^{\infty}\frac{e^{-t}}{t}dt$. Using the relation $\Sigma_{\alpha}^<(t_1,t_2)\big|_{t_1<t_2}=-\left[\Sigma_{\alpha}^<(t_1,t_2)\big|_{t_1>t_2}\right]^*$, the full expression of $\Sigma_{\alpha}^<(t_1,t_2)$ could be obtained.
The remaining components could be calculated by the relations,
\begin{align}
&\Sigma_{\alpha}^{>}(t_1,t_2) = \Sigma_{\alpha}^{<}(t_1,t_2)+ \Sigma_{\alpha}^{r}(t_1,t_2) - \Sigma_{\alpha}^{a}(t_1,t_2) ,  \notag \\
&\Sigma_{\alpha}^{t}(t_1,t_2) = \theta(t_1-t_2)\Sigma_{\alpha}^{>}(t_1-t_2)
+ \theta(t_2-t_1)\Sigma_{\alpha}^{<}(t_1-t_2) , \notag \\
&\Sigma_{\alpha}^{\bar{t}}(t_1,t_2) = \theta(t_2-t_1)\Sigma_{\alpha}^{>}(t_1-t_2) + \theta(t_1-t_2)\Sigma_{\alpha}^{<}(t_1-t_2) .
\end{align}
Note that the following relations hold
$\Sigma_{\alpha}^{++} = \Sigma_{\alpha}^{t}$, $\Sigma_{\alpha}^{+-} = -\Sigma_{\alpha}^{<}$,
$\Sigma_{\alpha}^{-+} = -\Sigma_{\alpha}^{>}$, and $\Sigma_{\alpha}^{--} = \Sigma_{\alpha}^{\bar{t}}$.



\end{document}